# Chemistry research in India in a global perspective – A scientometrics profile

Muthu Madhan,[ac] Subbiah Gunasekaran,[b] Rani M T,[c] Subbiah Arunachalam[c*] and T A Abinandanan[c]

[a]Azim Premji University, Bengaluru, India, 560100
[b]Central Electrochemical Research Institute, Karaikudi, India, 630006
[c]DST-Centre for Policy Research, Indian Institute of Science, Bengaluru, India, 560012.

Muthu Madhan [madhan.m@azimpremjifoundation.org]; Subbiah Gunasekaran [guna1970@gmail.com]; Rani M T [ranimt2503@gmail.com]; Subbiah Arunachalam [subbiah.arunachalam@gmail.com]; T A Abinandanan [abinand@iisc.ac.in]

*Author for correspondence

**Abstract**

We measure India's contribution to chemistry research in a global perspective. In the five years 2011-2015 Indian researchers have published 62,448 papers in 557 journals. In terms of % share, India (with 6.9% of the world's publications) is behind only China (25%) and USA (17%). But only 0.86% of papers from India are among the top 1% of the most highly cited papers of the world, compared to 4.86% of papers from Singapore, 2.65% of papers from USA, 2.09% of papers from China, 1.87% of papers from the UK, 1.71% of papers from South Korea and 1.6% of papers from Germany. Papers from India are cited 14.68 times on average compared to cites per paper of 45.34 for Singapore, 30.47 for USA, 23.12 for China, 26.51 for the UK, 21.77 for South Korea and 24.77 for Germany. Less than 39% of papers from India are found in quartile 1 (high impact factor) journals, compared to 53.6% for China and 53.8% for South Korea**.** Percent share of papers in quartile 1 journals from India is lower than that for the world for all of chemistry and for each one of the eight categories, viz. analytical, applied, inorganic & nuclear, medicinal, multidisciplinary, organic, physical and electrochemistry whether one considers data for the entire five-year period or for 2015 alone. About 20% of Indian chemistry papers are in collaboration with international coauthors. Researchers from only 160 Indian institutions have published at least 100 papers (compared to 362 in USA and 399 in China) and these include 67 state, 14 central and 11 private universities, 27 institutions under the Ministry of Human Resource Development, 20 CSIR laboratories, seven Department of Atomic Energy institutions, and seven Department of Science & Technology institutions. About 40% of all Indian chemistry papers have come from public universities. Only three Indian institutions, viz Bhabha Atomic Research Centre, Indian Institute of Science and Indian Institute of Chemical Technology, have published more than 2,000 papers. None of the Indian universities has performed as well as leading Asian universities. Amrita Vishwa Vidyapeetham, a small institution with less than 200 papers, has performed reasonably well.

**Introduction**

In an earlier paper we had shown that chemistry research in India, based on analysis of papers published in three multidisciplinary chemistry journals, viz. *Journal of the American Chemical Society*, *Angewandte Chemie International Edition* and *Chemical Communications*, is growing but much more slowly than in China.[1]

In this paper, we analyse India's contribution to chemistry research in a global perspective, particularly in comparison with China's, based on papers published in 644 journals. To this end, we have looked at the chemistry papers produced from countries that have published at least 4,000 papers during 2011-2015 and their citation performance. In all, 42 countries, including 14 from Asia, met the threshold. We have analysed the citation performance under four indicators, viz. cites per paper (CPP), papers in high impact factor journals (referred to as Quartile 1 or Q1 journals), and % of papers in the world's top 1% and top 10% most highly cited papers, as well as % of papers resulting from international collaboration. We have applied these indicators at three different levels, viz. country, field and institution. As the purpose of the study is to contribute to India's science policy, we looked at the performance of different types of institutions, viz. academic institutions such as state, central and private universities, and research institutions such as those under Council of Scientific & Industrial Research (CSIR), Department of Science and Technology (DST) and Department of Atomic Energy (DAE).

After commenting on why we think citation data can help in such studies, we show, using funding and manpower data at the institutional and country level, that China performs far better than India in chemistry research. China's performance owes it to sustained funding, ability to provide jobs and attract talent. Although in terms of number of papers India is No.3 in the world, it has not retained this position when it comes to citations. Indeed, it is behind many smaller countries in Asia and elsewhere.

**Methodology**

We gathered data on papers in chemistry published during 2011-2015 in many countries from *InCites*, the research analytics tool of Clarivate, during March 2019. We considered four document types, viz. article, letter, proceedings paper, and review.

In *InCites*, journals are classified using different schemas such as *Web of Science* (*WoS*) schema and *Essential Science Indicators* (*ESI*) schema. We used the *WoS* schema which classifies all of science into more than 250 categories.[2]

According to *InCites*, 907,934 papers in chemistry have appeared in 644 journals in the five years 2011-2015. These journals fall into eight categories, viz. analytical chemistry (comprising 127 journals), applied chemistry (89 journals), inorganic & nuclear chemistry (57 journals), medicinal chemistry (87 journals), multidisciplinary chemistry (222 journals), organic chemistry (81 journals), physical chemistry (180 journals), and electrochemistry (48 journals). Some journals may be classified under more than one category in *InCites*. For example, *Journal of Power Sources* is assigned four categories, viz. physical chemistry, multidisciplinary materials science, electrochemistry, and energy & fuels. When a journal is assigned a category, all papers in that journal will be counted under that category. Several categories make up a subject area. The size of different subject areas, in terms of the number of papers published, varies widely.

*InCites* reassigns individual articles in multidisciplinary journals such as *Nature* and *Science* to their most relevant subject area by using the information found in the cited references of each publication. "This reclassification process allows articles to be appropriately compared with articles of similar citation characteristics and topic focus".[3] If it is not possible to accurately reassign the publications (for example when the article does not have cited references) they are left as multidisciplinary.

According to *InCites*, during 2011-2015, authors from 191 countries published 900,200 papers in chemistry. This number is lower by 7,734 papers compared to the number computed from the 644 journals. That is because there may be papers which do not have the address field. *InCites* also provides data on number of papers from different regions/groupings such as European Union and OECD.

We analysed the performance of different countries using the following indicators: average citations per paper (CPP), % of papers in quartile 1(Q1) journals, % of papers in the top 1 percentile and the top 10 percentile of the most highly cited papers of the world. Also, we have considered % of papers involving international collaboration.

*Average citations per paper (CPP)*

'Citations per paper' (CPP) has been extensively used as a bibliometric indicator in research performance evaluation and can be applied at all organizational levels (author, institution, country/region, research field or journal). However, there are limitations to the indicator. For example, it ignores the total volume of research outputs.[4] The citability of papers varies from field to field and subfield to subfield due to several factors, such as the degree to which references from other fields are cited, and one should be careful about making inter-subfield and inter-field comparisons.

*The % of papers in Q1 journals*

*InCites* classifies journals in a category into four quartiles based on journal impact factors (JIF) provided in *Journal Citation Reports* published every year. The JIF quartile is the quotient of a journal's rank (X) in its category and the total number of journals in the category (Y), so that (X / Y) = Percentile rank Z.[5]

Q1: $0.0 < Z \leq 0.25$

Q2: $0.25 < Z \leq 0.5$

Q3: $0.5 < Z \leq 0.75$

Q4: $0.75 < Z$

A journal may be classified under different quartiles in different years as the quartile assigned is based on impact factor for the given year and its category. For example, *RSC Advances* had changed its quartile more than once during 2011-2017: Q4 in 2011, Q2 in 2012, Q1 in 2013 and 2014, and Q2 in 2015-2017. As we are dealing with a multi-year period each quartile will have more than 25% of all the journals. Also, "*InCites* displays the best quartile for journals that appear in multiple Web of Science Research Areas".[5]

*InCites* assigns papers from a multi-year collection to different quartiles based on the years in which the journal was classified under a quartile, which means only papers which were published in the years in which the journal was categorized under Q1 would be included under Q1. We have also considered India's contribution to Q1 journals in two other ways: papers in journals which are under Q1 in all five years, and papers in journals which are under Q1 in at least one of the five years.

*The % of papers in the top 1 percentile and the top 10 percentile most highly cited papers of the world*

'The % documents in top 1%' is the top one percent most highly cited documents in a given subject category, year and document type divided by the total number of documents in a given set of documents, displayed as a percentage.[6] 'The % documents in top 1%' may be an indicator of research excellence as only the most highly cited papers would make the top one percent in any level of aggregation (author, institution, national/international, field). It is best used with large datasets such as the accumulated publications of an institution, country or region and for a publication window of several years. 'The % documents in top 10%' is like the % documents in top 1% simply with a threshold of 10% instead of 1%.[6] OECD, among others, uses this indicator in their STI Scoreboard.[7]

*The % of international collaborations*

The measure we use for international collaborations is the number of papers that are coauthored by researchers from more than one country. The % share is the number of internationally coauthored papers divided by the total number of papers.[6] This may be considered as an indication of an institution's or author's ability to attract international collaborations.

We assigned the name of the city, the state, and the type for each institution. For publications from the laboratories under the Council of Scientific & Industrial Research, we standardized the names of institutions manually, as there were countless instances of name variations in author addresses.

**Analysis**

*Distribution of papers by region/group of nations*

In Table 1 we present data on the distribution of chemistry research in different regions/groups of nations. Almost 60% of papers (produced from countries in these groups) have at least one author from OECD countries and 49% of papers have at least one author from Asia Pacific. Papers from ASEAN, OECD, European Union (EU-15) and Nordic countries have recorded a higher CPP than the other regions or groups of nations. Africa and Latin America have recorded a low CPP.

OECD, European Union and Nordic countries have published at least 60% of their papers in Q1 journals.

In both % share in the top 1% and the top 10% of the most highly cited papers, ASEAN – it is just Singapore – is way ahead of the rest. In % of papers in the top 1% of the most highly cited papers, Asia Pacific and BRIC are behind ASEAN but ahead of OECD.

Nordic countries account for the largest % share of papers in international collaboration. Several regions – Africa, ASEAN, EU, Middle East, Latin America, and OECD - have a higher percent share of internationally coauthored papers than the world average.

*Distribution of papers by country*

Eight countries including India have published more than 40,000 papers in chemistry and 21 more than 10,000. (Table 2) China, which surpassed the USA in the number of papers published in 2009, leads the world with 25% of papers, followed by the US with 17%. India is a distant third with 6.9%. The big three of Europe, viz. Germany (6.8%), France (4.6%) and UK (4.4%), together accounted for 14.9%, and Japan 6.6%.

During 1997-2001, 168 countries had contributed 442,067 papers to the literature of chemistry indexed in journals covered by *InCites*. How could one explain the large increase in the number of countries contributing to chemistry research between 1997-2001 and 2011-2015? Several countries were not independent entities during 1997-2001; for example, Serbia, Montenegro and Kosovo were part of Yugoslavia and became independent much later. *InCites* shows that during 2011-2015, nine papers in chemistry were published from Kosovo, 53 papers from Montenegro and 3,206 papers from Serbia. Others like Bhutan, Laos and Central African Republic were very much there during 1997-2011 but *InCites* has not recorded any chemistry paper from those countries. Another reason could be the large increase in the number of journals publishing papers in chemistry research indexed by *WoS* - from 495 journals in 1997-2001 to 644 in 2011-2015.

In 1997-2001 India occupied the 10th position with 3.8% of world's papers in chemistry and China the 6th position with 5.8%, and USA was the world leader with 22.4%. Germany (9.5%), UK (7.1%), and France (6.8%) together accounted for 22.7%, and Japan 12.6%.

Between the two periods the production of journal literature in chemistry has more than doubled. The doubling was supported by the arrival of new mega journals some of which did not even levy article processing charges (e.g. *RSC Advances*) and journals devoted to narrow specialties as well as by already large journals publishing even larger number of papers (e.g. *Applied Surface Science* which published 4,181 papers in 1997-2001 and 10,894 papers in 2011-2015). Also, the number of journals publishing 2,000 papers or more grew from 57 in 1997-2001 to 116 in 2011-2015. But in the advanced countries, the growth in the number of papers published in chemistry was below the world average (104%): The number of papers having at least one author from USA grew by 55.4%. The number from Germany recorded a growth of 44.7%, from France 35.2%, from the UK 26.4% and from Japan 7.4%. In contrast, many Asian countries recorded phenomenal growth - China 779%, Singapore 560.8%; South Korea 338.8%, India 274%, and Hong Kong 184%. We have compared the number of papers in chemistry from India and China and a few selected advanced countries in two different periods, 1997-2001 and 2011-2015 in Fig. 1.

In terms of funds invested on R&D also, between 1999 and 2009 the U.S. share of global R&D dropped from 38 to 31 percent, whereas the share of ten countries of Asia put together grew the fastest - from 24 to 32 percent.[6] Clearly the US dominance in S&T is no longer as strong as it used to be.

*Distribution of papers by CPP*

Average CPP for papers published worldwide during 2011-2015 in chemistry is 20.57. Only 18 countries have a better average CPP relative to the world. Singapore (45.34) and Hong Kong (33.74) have recorded the highest CPP, followed by the USA (30.47), Switzerland (29.77) and the Netherlands (29.28).

Other Asian countries that have a CPP better than the world average are Israel (25.09), Saudi Arabia (24.36), China (23.12), and South Korea (21.77). Apart from these, there are seven more Asian countries in the top 42 countries, but their CPP is below the world average: Taiwan (19.26), Japan (19.00), Malaysia (16.04), India (14.68), Thailand (14.44), Iran (13.71), and Pakistan (11.12). India occupies the 10th position in Asia and the 28th position in the world.

In six of the eight categories, viz. multidisciplinary chemistry, physical chemistry, electrochemistry, organic chemistry, analytical chemistry and applied chemistry, China has a higher CPP than that of the world average for that category. For India, except for inorganic & nuclear chemistry, for all the others, the average CPP is lower than that for the world; however, papers in applied chemistry, electrochemistry and medicinal chemistry have CPP close to the world average. (Fig. 2)

*Distribution of papers in high impact factor journals*

In the five years 2011-2015, more than 7.1 million papers were published in all of science by researchers worldwide in 11,729 journals at a CPP of 14.95.

During the same five years 907,934 papers were published in 644 chemistry journals mostly published from 10 countries at a CPP of 20.45.

Journals published from four countries, viz. England (28.2%), USA (28.1%), the Netherlands (12%) and Germany (10.4%) account for more than 78% of the entire world's chemistry papers. Only 14 chemistry journals published from India have been indexed in *InCites* and these journals have published 5,005 chemistry papers (0.6%). Nine of the 644 chemistry journals have published more than 10,000 papers in the five years, and one of them more than 26,000. These nine journals account for 14.8% of all chemistry papers. A little more than 41% of papers in chemistry have been published in 43 journals each one of which have published at least 5,000 papers.

*Q1 journals are cited often*

There is a pecking order among journals based on citability over a long time; however, journals considered important by the community will rarely miss out being a Q1 journal. Papers published in Q1 journals - those having high impact factors - will normally be cited more often than those in the other three quartiles, and those published in quartile 4 journals – which have low impact factors – will on the average be cited the least.

*InCites* assigns papers from a multi-year collection to Q1 journals (considering only papers in the years in which a journal was in Q1). Of the more than 7.1 million papers in all of science published during 2011-2015, about 3.15 million were published in 4,199 Q1 journals and these had a CPP of 23;

about 1.7 million papers were published in 5,183 Q2 journals and these had a CPP of 10.4; about 1.1 million papers were published in 5,170 Q3 journals and these had a CPP of 6.41; and the rest of the papers published in 4,413 Q4 journals had a CPP of 3.5. (*InCites* data updated on 31 Jan 2019). Among papers published by a country as well as papers in a subject area, bulk of the citations go to Q1 journals. As we move from Q1 to Q4, there is a steep decline in CPP.

*Papers in Q1 journals by country of origin*

Among the 42 countries that have published at least 4,000 papers in chemistry, 24, including six from Asia, have published at least 53.82% (the world average) of their papers in Q1 journals. (Table 2). The Asian countries are Singapore (78.5% of them in Q1 journals), Hong Kong (79.91%), Israel (68.5%), Taiwan (65.24%), Japan (53.84%) and South Korea (53.82%).

In 2014 China crossed the world average in the percent share of papers in Q1 journals. Argentina is the only Latin American country which finds a place in this list (4,958 papers; 54.82 % of them in Q1 journals). India occupies the 34th position with 38.61% of its papers in Q1 journals. Many scientifically less advanced countries such as Brazil, Mexico, Thailand, South Africa and Hungary are above India. The distribution of papers from many Asian countries into different quartiles shows the relative position of India. (Fig. 3) While India has recorded only a modest increase in % of papers in Q1 journals between 2001-2005 and 2011-2015, Taiwan, Saudi Arabia and Mainland China have made substantial gains.

We compare the growth in the overall number of papers in chemistry published from India and China during the five years as well as the growth in the number of papers in Q1 journals. (Fig. 4) The number of papers published by China grew from 34,426 in 2011 to 57,780 in 2015 for a compound annual growth rate (CAGR) of 13.82%, and the number of papers published by China in Q1 journals grew from 17,121 in 2011 to 30,073 in 2015 for a CAGR of 15.12%. Q1 journals account for a higher rate of growth than all journals by about 1.3%. The number of papers published by India grew from 10,641 in 2011 to 14,593 in 2015 for a CAGR of 8.2%, and the number of papers published by India in Q1 journals grew from 3,572 (33.56%) in 2011 to 4,972 (34.08%) in 2015 for a CAGR of 8.6%. Q1 journals account for a higher rate of growth than all journals by about 0.4%, much less than in China.

We notice a spike in the % of papers from India and China in Q1 journals in 2013 and 2014 (Fig. 4) and we attribute it to papers published in *RSC Advances* which was classified under Q1 in those two years. Indeed, *RSC Advances* was the most often used chemistry journal for India from 2013 to 2015 and for China in 2014 and 2015. In the two years 2013 and 2014, China (5,183 papers) and India (2,140 papers) accounted for more than 62% of the 11,725 papers published in the journal. The 577 papers from India in *RSC Advances* in 2013 helped India's share of Q1 papers increase by more than 2.5%, and the 1,307 papers from China helped its share increase by more than 1.5%. In 2014, publications in *RSC Advances* helped improve China's share in Q1 journals by more than 3% and India's share by over 6.4%. The corresponding figures for the world are 0.9% in 2013 and 1.93% in 2014.

*Q1 journals in all 5 years*

In Table 3 we have considered all the papers published in journals that were Q1 in all the years or in any one of the years and journals which were not Q1 in any of the years. Of the 644 journals which had published 907,934 papers in chemistry during 2011-2015, 136 were classified under Q1 in each one of the five years and these had published 393,649 papers (CPP of 33.1); and 96 journals were classified under Q1 in at least one of the five years and these had published 158,713 papers (CPP 16.8). That some of these 96 journals were not in Q1 in initial years is a mere artifact, and not because these journals were not quality journals, e.g. *ACS Catalysis* that commenced publication in 2011 was categorized as a Q4 journal but from 2012 onwards it is a Q1 journal, and *Journal of Materials Chemistry A* entered the database in 2013, the year *Journal of Materials Chemistry* was split into three parts, as a Q4 journal but moved to Q1 the very next year. All the other 412 journals have not been in Q1 in any of the five years and together had contributed 355,572 papers (CPP 8).

Journals which are in Q1 in all five years (accounting for 43.3% of the world's papers, 43.1% of papers from China and 27.4% of papers from India) have a much higher CPP than other journals for the world or for individual countries: 33.1 for the world, 38.3 for China, and 26.3 for India. The average CPP of Indian papers in Q1 journals is considerably lower and that of the Chinese papers considerably higher than that of the world, whichever way Q1 is defined. As pointed out by Arunachalam and Manorama as early as in 1988, Indian papers published in high impact factor journals, in general, do not get cited to the expected level.[8]

*Papers in Q1 journals by subfield*

Percent share of papers in Q1 journals in electrochemistry (68.9%), physical chemistry (63.9%) and applied chemistry (60.3%) was higher than the world average for all of chemistry (53.5%).

If we consider the % share in just one year, 2015, 50.2% of all world's papers in chemistry were in Q1 journals, with electrochemistry (65.1%), applied chemistry (65%), physical chemistry (61.9%) and analytical chemistry (54.6%) having a higher share than the world average for all of chemistry.

Percent share of papers in Q1 journals from China in four subfields, viz. electrochemistry, analytical chemistry, applied chemistry and organic chemistry was higher than that for the world if we consider data for the five years 2011-2015 (Fig. 5). However, if we consider only papers published in 2015, apart from these three subfields China has a higher share than that for the world in two other subfields, viz. analytical chemistry and applied chemistry. Percent share of papers in Q1 journals from India is lower than that of the world for all of chemistry and for each one of the eight categories.

*Percent of papers in lower tier journals*

Worldwide 9.5% of papers were published in Q4 journals, counting only papers in the years in which a journal was in Q4. In 14 countries, though, a larger percent of papers was published in Q4 journals, e.g. Russia (53.36%), Ukraine (40.60%), Pakistan (26.12%), and Romania (21.5%). India is in the 9[th] position with 13.55% of papers in Q4 journals and China in the 12[th] position with 9.95% of papers in Q4 journals.

*Percent of papers in top 1 and top 10 percentile of highly cited papers*

In chemistry 1.45% of papers published worldwide are in the top 1 percentile of the most highly cited papers. Among the countries that have published at least 4,000 papers, 15 have published more than 1.46% of their papers in the top 1 percentile of highly cited papers. Six Asian countries, viz. Singapore (4.86%), Hong Kong (3.33%), Saudi Arabia (2.95%), China (2.09%), South Korea (1.71%) and Israel (1.59%) are among these countries. India (0.86%) is at the 29th position. Smaller countries such as Taiwan (1.19%), Malaysia (1.12%), Thailand (0.98%) and Turkey (0.88%) are above India.

The number of papers in the top 1 percentile of the most highly cited papers for the world grew from 2,478 papers in 2011 to 2847 in 2015 for a CAGR of 3.53%. For China the number grew from 721 in 2011 to 1190 in 2015 for a CAGR of 13.35%. For India the number grew from 97 in 2011 to 115 in 2015 for CAGR of 4.35%.

In chemistry research worldwide 13.01% of papers are in the top 10 percentile of the most highly cited papers, and only 18 countries among those that have published at least 4,000 papers have a larger percent share of papers in the top 10 percentile. Singapore (28.93%), Hong Kong (23.86%) and USA (19.40%) are the top three countries. Several Asian countries have a score better than the world average: Saudi Arabia (16.91%), China (16.86%), Israel (14.84%), and South Korea (13.95%). India occupies the 29th rank with 9.43%, and Taiwan (11.48%), Malaysia (10.66%), Japan (9.86%), Iran (9.52%) and Thailand (9.65%) are above India.

The number of papers in the top 10 percentile of the most highly cited papers for the world grew from 21,887 in 2011 to 26,130 in 2015 for a CAGR of 4.53%. For China the number grew from 5,757 in 2011 to 10,080 in 2015 for a CAGR of 15.03%. For India the number grew from 986 in 2011 to 1305 in 2015 for a CAGR of 7.26%.

The %share in the top 1 and 10 percentile of the most highly cited papers is higher in four fields, viz. physical chemistry (1.87% and 15.84%), multidisciplinary chemistry (1.8% and 14.49%), applied chemistry (1.51% and 14.36%), and electrochemistry (1.45% and 13.43%), than in all of chemistry.

In percent share of the top 1 percentile of the most highly cited papers China has performed better than the World average in each subfield except medicinal chemistry. China has done well in Analytical chemistry (1.9%), physical chemistry (2.95%), organic chemistry (1.76%), electrochemistry (2.22%) and multidisciplinary chemistry (2.5%). India's share in the most highly cited one percentile papers is less than the world in all categories other than Inorganic & nuclear chemistry. (Fig. 6) But, if we consider percent share of papers in the top 10 percentile of the most highly cited papers, then India has done better than the world average in three categories, viz. Electrochemistry, Medicinal chemistry and Inorganic & nuclear chemistry. China has a higher percent share of papers among the top 10 percentile of the most highly cited papers in each of the eight categories than the world. (Fig. 7)

**Institutions performing chemistry research**

We must begin with a caveat. Field weighted percentile indictors work well only with large samples, e.g. country-level data or data covering large institutions.[9, 10] While evaluating institutions which produce a small number of papers, one must be cautious in using these indicators. Many Indian institutions fall under this category. Even so we use these indicators as these can help identify institutions that produce better cited papers. Our intention is not to compare institutions.

In all, as on 3 April 2019, according to *InCites*, about 8,740 institutions from 191 countries, including about 760 from India, 910 from China, 250 from the UK and 1,300 from the USA have published at least one paper in chemistry during 2011-2015. These numbers are not exact as apart from individual institutions under a large group the group is also listed sometimes, e.g. individual IITs and the IIT system or individual CSIR laboratories and CSIR.

160 Indian institutions, 399 Chinese institutions, 71 institutions from the UK and 362 institutions from the US have published at least 100 papers during this period. Of these, 21 Indian institutions, 73 Chinese institutions, 54 in the UK and 263 in the USA have published at least 60% of their papers in Q1 journals. If we raise the bar to 75% of papers in Q1 journals, the numbers drop to 95 institutions in the US, 11 in the UK, nine in China, and only four from India. A larger percent of institutions in the USA and UK contribute in Q1 journals than institutions in China and India. This is to be expected, as both the UK and the USA have a long tradition of doing modern chemistry research.

The average citation impact varies considerably among institutions. Of the 362 institutions in the US that have published at least 100 papers in chemistry, 197 have a CPP ≥ 25 and 350 have a CPP ≥ 15. Of the 399 Chinese institutions that have published not less than 100 papers, 72 have a CPP ≥ 25 and 242 have a CPP ≥ 15. Of the 71 institutions in the UK that have published 100 or more papers, 30 institutions have a CPP ≥ 25 and 70 have a CPP ≥ 15. Only six of the 167 Indian institutions that have published not less than hundred papers have a CPP ≥ 25 and 74 have a CPP ≥ 15 as on 3 April 2019.

At least 5% of the papers from 25 institutions in the US, 12 institutions in China, and one in the UK are in the top 1% of the most highly cited papers of the world. No institution from India has at least 5% of its papers in the top 1% of the most highly cited papers. With 4.79% of papers in the top 1% of the most highly cited papers Amrita Vishwa Vidyapeetham comes close, but Amrita has produced only 188 papers of which just eight papers are in the top 1 percentile. At least 25% of papers from 51 US institutions, 18 Chinese institutions and one from UK and one from India (Indian Institute of Science Education & Research, Bhopal) are in the top 10% of the most highly cited papers of the world.

**Indian institutions**

India has published 62,448 papers in chemistry during the five years 2011-2015, but the publishers of *InCites* have assigned correct institutional addresses to only 57,115 papers from 761 institutions as on 3 April 2018. They are working on assigning addresses to the remaining institutions. (Private communication, Wesley Keller, 4 Dec 2018).

In this analysis, we have considered only papers from 160 institutions which have published at least 100 papers, the total amounting to 52,114 papers (91.2%) for which addresses have been assigned. These include 92 universities (state, central and private), 27 MHRD institutions, 20 CSIR laboratories, seven DAE institutions, seven DST institutions and seven other entities.

*State and Central Universities*

Only 81 Indian universities have published at least 100 papers in the five years, and they have published 25,119 papers (or 40.2% of all Indian chemistry papers) at an average CPP of 13.6. Of these 31.2% of papers have appeared in Q1 journals, 0.73% of papers are in the top one percentile of the world's most highly cited papers and about 9% in top 10 percentile. About 20% of them are internationally coauthored.

Five universities, viz. Jadavpur University, Delhi University, Banaras Hindu University, University of Hyderabad and University of Calcutta, have published at least 1,000 papers in the five years, and 14 have published at least 500 papers. (Table 4)

Just one university, viz. Shivaji University (21.63), has a CPP higher than the world average, with three others, viz. Guru Nanak Dev University (19.53), Jamia Milia Islamia (19.51) and Madurai Kamaraj University (19.06) coming close.

None of the 81 public universities has a percent share of papers in Q1 journals even close to that of the world (53.8%); however, Jawaharlal Nehru University has 50.5% of its papers in Q1 journals and five others, viz. Guru Nanak Dev University, Tezpur University, University of Hyderabad, Shivaji University and Pondicherry University have more than 45% of their papers in Q1 journals. Eleven universities in all have more than 40 % of papers in Q1 journals.

Twelve universities have a higher percent of papers in the top one percentile most highly cited papers than the world average. Himachal Pradesh University, Guru Nanak Dev University, Visva-Bharati, University of Mumbai, and Bharathidasan University have more than 2% of papers in the top 1 percentile, three times the overall %share of all state and central universities. Guru Nanak Dev University, University of Hyderabad, Banaras Hindu University and Jadavpur University have published at least 10 papers in the top 1 percentile.

Twelve universities have published a larger percent of papers in the top 10% most cited papers than the world average. Only four universities viz. Shivaji University (19.40%), Bharathiar University (16.92%), Gandhigram Rural Institute (16.32%), and Mahatma Gandhi University (15.66%) have published at least 15% of all their papers published in the top 10% most highly cited papers.

At least 25% of the papers published from each one of 18 universities have an author from countries other than India.

*Private universities*

Eleven private universities have published at least 100 papers in chemistry journals in the five years and together they have published 2,648 papers at an average CPP of 13.81. About 32% of these papers have appeared in Q1 journals. 0.98% of these papers are among the most highly cited 1% of papers in the world and 9.78% are among the most cited 10%. About 23% of these papers are internationally coauthored.

Amrita Vishwa Vidyapeetham University has the highest CPP (31.21) among all Indian institutions, and 58% of its papers are in Q1 journals, the highest for Indian universities; 4.8% and 23.4% of its papers are in the top 1% and top 10% of the most highly cited papers of the world. Vellore Institute of Technology and Birla Institute of Technology and Science (BITS), Pilani, have published more than 500 papers each. Five other universities, viz., SRM University, Birla Institute of Technology Mesra, Manipal University, BITS Pilani, and Loyola College, Chennai, have published at least 30% of their papers in Q1 journals. More than one third of the papers from two private universities (Karunya Institute of Technology, 38%, and Amrita Vishwa Vidyapeetham, 35%) are internationally coauthored.

*MHRD institutions*

The 27 MHRD institutions that have published at least 100 papers in chemistry include Indian Institute of Science (IISc), 13 Indian Institutes of Technology (IITs), five Indian Institutes of Science Education & Research (IISERs), seven National Institutes of Technology (NITs), and Indian Institute of Engineering Science and Technology, Shibpur.

Indian Institute of Science (IISc) has published more than 2,000 papers, and four IITs (Bombay, Kharagpur, Madras and Kanpur) have published more than 1,000 papers each. The 27 MHRD institutions together have published 13,530 papers at a CPP of 18.88. Seven institutions have an average CPP of above 20: IISER, Thiruvananthapuram (27.92), NIT Rourkela (26.35), IIT Roorkee (25.39), IISER Pune (24), IISER Bhopal 23.34 and IISc Bangalore 22.75. [Table 6] A few highly cited papers have contributed to the relatively high CPP of some institutions, especially those which have published fewer papers. For example, a paper from NIT Rourkela and another from IISc, Bangalore, have been cited more than 1,000 times, and two papers each from IISc and NIT Rourkela, and one from IIT, Roorkee, have been cited more than 500 times; 48 papers from IISc, 27 from IIT Roorkee, 13 from IISER Pune, six from IISER Thirvananthapuram and four from IISER Bhopal have been cited not less than 100 times.

In all 54.08% of papers published by 27 MHRD institutions are in Q1 journals. Thirteen MHRD institutions have published more than 50% of their papers in Q1 journals. Three of them, viz. IISER Thiruvananthapuram (78.83%), IISER Pune (75.87%), and IISER Bhopal (74.17%) have published more than 70% of their papers in Q1 journals. Collectively IISERs became the fourth-ranked Indian organization in chemistry in the Global 500 in 2015 in *Nature Index*[11] and further improved to second rank after the IITs as per data provided in *Nature Index* for the period 1 November 2017 - 31 October 2018. < https://www.natureindex.com/ > IISERs have performed well not only in chemistry, but in other disciplines as well, says a biologist.[12]

Of the 13, 530 papers published from the 27 MHRD institutions, 200 (or 1.48%) are in the top 1% and 1,788 (or 13.22%) are in the top 10% most highly cited papers of the world. Of the 234 papers published from the Indian Institute of Technology, Dhanbad, 11 (4.7%) are among the top 1% most highly cited papers. Seven other MHRD institutions have higher than 2% of their papers in the world's top 1% most highly cited papers. Two institutions, viz. IISER Bhopal and IISER Thiruvananthapuram, have more than 20% of their papers in the top 10% of the most highly cited papers of the world, 12 institutions have more than 15% of their papers in the top 10 percentile of the world's most highly cited papers.

Five IITs (Roorkee, Ropar, Bombay, Kanpur and Varanasi) and two NITs (Allahabad and Tiruchirapalli) have coauthored more than 25% of their papers with researchers from other countries.

*CSIR institutions*

CSIR laboratories have published 10,250 papers in the five years at an average CPP of 16.92, and about 48% of these papers have appeared in Q1 journals. A meagre 0.9% of them are in the top 1 percentile of the most highly cited papers of the world and 10.68% in the top 10 percentile. Less than 16% are internationally coauthored. Twenty CSIR laboratories have published 100 papers in the five years, including eight with more than 500 papers and two with more than 1,000 papers. (Table 7)

The average CPP of papers from National Physical Laboratory (NPL) and National Institute for Interdisciplinary Science and Technology (NIIST) is above that of the world, and Institute of Minerals and Materials Technology (IMMT) and National Chemical Laboratory (NCL) come close to that. Nine laboratories, viz. NPL, IMMT, Central Glass and Ceramic Research Institute (CGCRI) NIIST, Central Electrochemical Research Institute (CECRI), NCL, Indian Institute of Petroleum (IIP), and Central Food Technological Research Institute (CFTRI) have a higher % share of their papers in Q1 journals than the world average. NPL, IMMT, NIIST and CSMCRI have at least 1.5% of their papers in the top 1 percentile of the most highly cited papers of the world. 18 papers from NCL, 11 each from Indian Institute of Chemical Technology (IICT) and NPL, 10 papers from CSMCRI, 7 papers from CECRI and 4 papers each from IMMT and CDRI are in the top 1 percentile. NPL, IMMT and CGCRI have a larger percent share of their papers in the top 10 percentile of the world's most highly cited papers than the world average. More than 20% of papers from six CSIR laboratories, with NEERI being the highest (about 50%), have an international coauthor.

*DAE institutions*

Seven DAE institutions have contributed 3,582 papers, with 51.59% of them in Q1 journals at an average CPP of 13.23. More than 75% of papers from Institute of Physics (IOP), Bhubaneswar and more than 60% papers from Raja Ramanna Centre (RRC), Indore, Tata Institute of Fundamental Research (TIFR), Mumbai, and National Institute of Science Education & Research (NISER) Bhubaneswar are in Q1 journals. More than 30% of papers from both TIFR and NISER have international collaborators. Bhabha Atomic Research Centre (BARC) is only one of three institutions in India to have published more than 2,000 papers, the other two being IISc and IICT.

*DST institutions*

Seven DST institutions have published at least 100 papers and together have contributed 2,861 papers, with Indian Association for the Cultivation of Science (IACS) contributing more than 1,400 papers, with average CPP 20.30. The CPP of Jawaharlal Nehru Centre for Advanced Scientific Research (JNCASR)(29%) and Sree Chitra Tirunal Institute for Medical Sciences Technology (20.87%) are greater than that of the world average. 61% of all papers from these seven DST institutions are in Q1 journals. More than 70% of papers from Sree Chitra Tirunal and JNCASR are published in Q1 journals and more than 2% are in the top 1% most highly cited papers.

*Other institutions*

Apart from those listed above, there are two UGC institutions, viz. UGC-DAE Consortium for Scientific Research (196 papers, CPP 11.27) and Inter-University Accelerator Centre (140 papers, CPP 9.88), the Defense Research Development Organisation system (899 papers, CPP 11.67), the Department of Biotechnology system (205 papers, CPP 13.58), Indian Council of Agricultural Research (387 papers, CPP 9.48), National Institute of Pharmaceutical Education and Research, Mohali (649 papers, CPP 14.92), and Advanced Centre of Research in High Energy Materials, Hyderabad (151 papers, CPP 13.74).

**International collaboration**

Indian chemists have collaborated with researchers from 115 countries to publish 12,403 papers (or 20 % of all papers from India) during the five years at an average CPP of 20.2 about 47.68% of these papers are in Q1 journals; 1.48% of them are in the top 1 percentile of highly cited papers and 14.35% are in the top 10 percentile. In contrast, papers without international collaboration have an average CPP of 12.58 and only 36.4% of them are in Q1 journals; 0.7% and 8.2% are in the top 1 and the top 10 percentile of the most highly cited papers. That collaboration helps improve citation impact and international collaboration even more so is well known.

The main collaborating countries are USA (2436 papers; CPP 23.62%), South Korea (1481 papers; CPP 21.02%, and Germany (1099 papers; CPP 19.37%). Of the Indo-German papers, more than 61% are in Q1 journals. India has collaborated with 23 South African institutions to publish 385 papers, of which 3.4% are in the top 1% of the world's most highly cited papers. India has collaborated with 24 Saudi Arabian institutions to publish 868 papers, of which 3.6% are in the top 1% of the most highly cited. India has collaborated with 142 institutions in China to produce 363 papers and these have an average CPP of 26.2.

Indian researchers have collaborated with 79 foreign institutions in at least 50 papers in the five years. CNRS, France, is the leading collaborating institution with 439 papers (CPP 22.58). Other leading collaborators are King Saud University (426 papers, CPP 16.56), King Abdul Aziz University (196 papers, CPP 23.2), Howard University, USA (153 papers, CPP 12.67) and Universiti Sains Malaysia (144 papers, CPP 14.78).

The 53 papers published in collaboration with Rice University have a CPP of 83.66. The 125 papers published in collaboration with King Fahd University of Petroleum & Minerals have a CPP of 71.25. The 85 Indian papers with CSIC in Spain have a CPP of 50.41.

At least 70% of papers resulting from collaboration with 15 institutions are in Q1 journals. These include University of Stuttgart (53 papers, 88.68% in Q1), Rice University (53 papers, 83.66% in Q1), University of Manchester (67 papers, 80.6% in Q1), and US Department of Energy (112 papers, 80.35% in Q1).

Though the number of papers is small, at least 5% of papers resulting from collaboration with 12 overseas institutions are in the top 1% of the most highly cited papers. King Fahd University of Petroleum & Minerals, Saudi Arabia has published 125 papers with Indian researchers, and of these 16.8% and 46.4% are in the top 1% and 10% of world's most highly cited papers. University of Johannesburg has published 68 papers with Indian coauthors and of these 8.82% and 32.35% are in the top 1% and top 10%. Islamic Azad University, Iran, has published 59 papers with Indian authors of which 5.1% and 30.5% are in the top 1% and the top 10%. Texas A&M University has published 71 papers with coauthors from India of which 7% and 33.8% are in the top 1% and the top 10%.

**Discussion**

*Is citation analysis a valid tool?*

Many researchers have reservations about the use of citation-based indicators. That is because of the careless use of quantitative citation data, a practice Garfield had dissociated right from the beginning.[13]

The uninformed use of indicators such as average and cumulative impact factors and the arbitrary criteria stipulated by some institutions and countries for the award of student fellowships and selection and promotion of faculty have made it difficult to distinguish good science from the bad and the indifferent.[14] While blind dependence on bibliometrics can lead to unethical practices in scholarly communication and the reward system in science, selective use of bibliometric measures can assist in enlightening multi-pronged assessments.[15]  Arguments that bibliometrics should never be used in the assessment of academic merit are erroneous.[16] As pointed out by Krull et al. , "evaluation practices are considerably influenced by the national and/or organizational context", and "successful evaluations heavily depend on the choice of the methods."[17]

Indeed, ever since National Science Foundation started compiling its biennial *Science & Engineering Indicators* in 1972, it has been using citation data as an indicator of the perceived usefulness of research results to further advance the state of knowledge.[18] According to Goroff of the Alfred P Sloan Foundation, "The productivity of science will depend on how well we can use bibliometrics" and "whether bibliometry is fit for a purpose depends on what the purpose is." The Sloan Foundation uses bibliometrics as a key tool in selection of early career fellows. The results are there for all to see: Of about 5,680 fellows selected since its inception in 1955 till the end of 2018, 47 have gone on to win Nobel Prizes, 16 Fields Medals, 69 National Medals of Science, 17 John Bates Clark Medals, and many others other distinguished awards.[19]

Stang, editor of *Journal of the American Chemical Society* (JACS), marked the journal's completion of 124[h] years with the publication of a list of the 125 all-time most-cited papers and the journal's impact factor over a 12-year period.[20]  Commenting on Stang's editorial Balaram observed that breakthrough papers could be identified by the number of citations.[21] According to Campbell, editor of *Nature*, "A higher number of citations is an indicator of significance, and usually of positive significance," although not a measure of excellence.[22] A few years ago, *American Economic Review* appointed a committee of six experts including two Nobel Laureates to identify the twenty most admirable and important papers published in the first hundred years of the journal and, as a first step, the committee used citation data and JSTOR search data to arrive at a large group of eligibles.[23] Out of this large set of papers, each member chose twenty and there was an encouraging unanimity or near-unanimity, especially about the 15 leading candidates, clearly showing peer review and citation-based evaluation converge.

According to Golitz, former editor of *Angewante Chemie*, often high impact factors go hand in hand with researchers' perception of quality and "papers in good journals are not better than those in other journals because they generate later more citations but rather because editors and referees read them carefully and made decisions based on their careful evaluations. Readers want to be sure that the manuscripts selected for publication in a journal were critically evaluated." [Personal communication,  8 May 2016]. Indeed, as Tragg says, removing journal level metrics from evaluation in favour of article level metrics does not negate the influence of journals.[24] One reason for the relatively high esteem of Indian Institutes of Science Education and Research, according to Shashidara, is that researchers there publish "in top journals in their fields of research; the kind of journals wherein referees dig deeper and deeper and, thereby, help improve the quality of publications."[12]

*Chemistry research in India relative to other Asian countries*

India's rank in publication output improves with an increase in the number of journals, but if we consider only publications in high-quality journals the rank falls.[1] India's chemistry research output is only second to China in Asia and third in the world. If we consider the number of papers in Q1 journals only (24,111), India's rank slides to the eighth, and if we consider the number of papers in the top 1 percentile of the most highly cited papers (537) India occupies the seventh rank. In all these three indicators, China occupies the first rank: 120,841 papers in Q1 journals, 4711 papers in the list of most highly cited papers.

India's rank falls further when it comes to its % share of chemistry papers in Q1 journals. India ranks 34th among the 42 countries that have published at least 4,000 papers during 2011-2015 and 10th among the 13 Asian countries. It is not just in the number or % of papers published in quality journals we observe such a decline in rank. In the top 1% of the most highly cited papers, with 0.86% India is 27th and there are 10 Asian countries above India, and in the top 10% of the most highly cited papers, with 9.44% India is 29th and there are 11 Asian countries above India.

The performance of chemistry research in leading universities in Asia, including Tsinghua and Beijing Universities, is far better than that of any Indian institution of higher learning or research. For example, Nanyang Technological University had published over 5,100 papers during 2011-2015 at a CPP of 51 with 81% of its papers in Q1 journals and 6.5% of its papers in the top 1 % of the world's most highly cited papers. Korea Advanced Institute of Science & Technology had published over 2,980 papers during 2011-2015 at a CPP of 37.83 with 78% of its papers in Q1 journals and 4.15% of its papers in the top 1 % of the world's most highly cited papers. In contrast, IISc had published 2,079 papers at a CPP of 22.75 with 62.82% of its papers in Q1 journals and 2.12% of papers in the top 1% most highly cited papers.

As seen from *InCites*, in the two decades 1999-2018, chemistry research in India and China grew faster than the world, but the growth is much more rapid in China. Total number of papers published all over the world grew by a factor of two, whereas it grew by 4.9 times in India and 16.2 times in China. Number of papers in Q1 journals grew by 8.2 times in India and 46 times in China as against 1.8 times for the world. Share of papers in the top 1 percentile of the most highly cited papers grew by 9.6 times in India and 75.6 times in China as against 2.4 times for the world.

In terms of the number of papers, during the same period the percent share for China grew from 5.9% to 34.9%, and for India it grew from 3.8% to 7.1%.  In terms of citations, the percent share for China grew from 3.5% to 46%, and for India it grew from 2.4% to 5.6%.  Also, % share of citations to papers from India was less than the % share of publications for each of the years 1999-2018, whereas it was less than the % share of publications in China to begin with and up to 2009 but changed direction in 2010 and started climbing.

The more stringent the criterion of evaluation one would expect lower would be the rating. Indeed, it is the case for India as we see from its % share of world's chemistry papers and its % share in papers published in Q1 journals and % share of papers which are in the top 1 percentile of the most highly cited papers (Fig. 8). However, in the case of China, one sees just the opposite trend (Fig. 9). As we move from the total output of the country to papers published in Q1 journals and the number of papers which are in the top 1 percentile of the most highly cited papers the share of papers from China is increasing. China's % share of papers in Q1 journals (28.5%) crossed it's % share of papers (27.4%) in 2014. China's % share of the world's most highly cited papers (17%) crossed the % share of papers (16.1%) in 2007.

As per *Nature Index* released on 12 December 2019, China is the biggest producer of high-quality research in chemistry, ahead of the United States. China's chemistry output is almost double the collective share of Japan, South Korea, and India.[25]

*What is behind China's performance?*

The enormous manpower deployed in China and the abundant and consistent investment made are key factors behind China's rapid growth.[26] By the end of 2017, China had 6.2 million employees in R&D.[27] There were about 0.41 million R&D and auxiliary personnel (FTE) in India as of 2015[28]

China is one of the leading producers of doctorates in science and engineering. In 2018, Chinese universities awarded 58,032 Ph.Ds of which 20,492 were in engineering, 12,208 in science, 9,561 in medicine and 2,654 in agriculture.[29] Female candidates accounted for 39% of all the Ph Ds awarded in 2018.[29] In 2017, Indian universities awarded 34,400 Ph.Ds, of which 8,880 were in science and 4,907 were in engineering and technology. Female candidates accounted for 41.3%.[30]

Both absolute amount of investment and rate of growth were higher in China in the past 25 years. China's R&D spending surged by an average of 20.3% annually between 1992 and 2017, with the amount for 2017 reaching 1.76 trillion Yuan (about $257 billion or 2.12% of China's GDP).[31] Businesses—including foreign-owned corporations— accounted for 1.36 trillion Yuan ($196.4 billion).[32] China plans to increase its annual per capita spending on R&D to 500,000 Yuan by 2020, up from 370,000 Yuan in 2014.[33] However, we do not have data for investments in R&D in Chemistry alone. Chinese universities are now well endowed and comparable to some top US institutions. For example, the annual expenditure of Tsinghua University in 2016 (US$3.57 billion at purchasing power parity) was higher than that of MIT and Yale University and that of Peking University ($2.45 billion) was higher than that of Caltech.[34]

R&D investment in India grew by a factor of 11.4 times between 1995-96 (Rs 7483.88 crore) and 2014-15 (Rs 85326.1 crore) at a CAGR of 13.67%.[35] According to data provided by DST, 38% of 2014-15 funding for R&D came from private sector.[36] None of the leading Indian universities spends even a tiny fraction of what the Chinese universities do. As seen from the Expenditure statements given in the respective annual reports, IISc spent $0.54 billion (at purchasing power parity), and IIT Bombay $0.47 bn, IIT Kharagpur $0.39 billion, IIT Kanpur $0.27 bn and IIT Madras 0.24 billion in the fiscal year 2017-2018.

Chinese universities have the advantage of scale. They have large numbers of students and faculty (those involved in teaching or research). THE World University Ranking 2019 has listed 1258 universities of which 55 are from China and 44 are from India. The top four Chinese Universities , viz. Tsinghua (36,912 students; 11% international) Peking (42,547; 17%) Zhegiang (36,031;18%) and Fudan(34,393;11%) have a much larger number of students and international students than the top four Indian institutions, viz. Indian Institute of Science (4,071 students, 1% international), Indian Institute of Technology Bombay (10,001; 1%), Indian Institute of Technology Kharagpur (9,397; 0%) and Indian Institute of Technology Roorkee(7,925; 2%). The Chinese universities also have a much larger number of faculty: Tsinghua (3,102), Peking (5,188), Zhegiang (3,306), Fudan (2,819), IISc (433), IITB (585), IITKh (676), IITR (455).

China's calculated plan to attract established scientists and high-tech entrepreneurs as well as highly productive postdocs to return to China is paying rich dividends. Between 2008 and 2018, more than 3,500 young scientists have returned to China.[37] India has also been trying to attract Indian researchers from abroad through programmes such as Ramanujan Fellowship, Ramalingaswamy Re-entry Fellowship, Visiting Advanced Joint Research Faculty Scheme (VAJRA) and INSPIRE Faculty Scheme. As a result, about 890 scientists have returned to India during 2007-2017.[38] Unfortunately, the INSPIRE scheme has run into some rough weather already with only a few Fellows getting jobs at host institutions.[39]

*Does funding help?*

Evidence on the effect of funding on research productivity and impact is not uniform. However, the few papers that we have come across show funding leads to increase in the number of publications. Based on a study of twenty years of federal and non-federal funding of chemistry research in the US, Rosenbloom et al. show that research funding increased publications and citations.[40] Based on all applications received for standard research grants from National Institutes of Health during the 20-year period 1980-2000, Jacob and Lefgren show that the grants usually led to only a marginal increase (approx. 7%) in the number of publications in the next five years.[41] Examining huge volume of publications resulting from about 25 years of research in more than 90 research universities in the light of spending, income and endowment data, Whalley and Hicks conclude that spending has substantial positive effect on the number of papers produced but not their impact.[42]

Verma believes that considering the level of funding and 'poor infrastructure' provided, Indian chemists are really doing well.[43]

Of the chemistry papers from India that have acknowledged funding, about 34,700 were supported by more than 350 funding agencies, with CSIR, DST and UGC accounting for 89% of the papers funded. More than 50% of DST-funded papers have appeared in Q1 journals and 1.22% of them are in the top 1% of the most highly cited papers and 12.89% are in the top 10% most highly cited papers. Also acknowledged are NSF, NIH and DFG, mostly in papers written in collaboration with US/German scientists. A higher percent of these papers has appeared in Q1 journals than those funded by Indian agencies only.

*What could India do?*

The critical factor that enabled the IISERs to emerge as leading scientific institutions in India was the care with which people were identified for leadership roles and the academic freedom given to them.[44] Each one of these IISERs has published at least 100 papers in chemistry journals during 2011-2015 and in citation-based indicators four of them have outperformed IISc and are approaching many Western universities. In 2018, *Journal of Physical Chemistry* has recognized 82 scientists, including five from India, as important to the future of physical chemistry, of which at least three are from IISERs[45] With only ten years since the first IISER was set up, it is still early days, and if they must grow and thrive, long term funding and autonomy must be ensured.

It is a matter for concern that many central universities set up around the same time as IISERs have not been able to perform as well as IISERs (It is interesting that Zare[46] also makes a similar observation in his response to Satyamurthy's Current Science editorial[47]). Of course, these universities do not enjoy the same level of funding and freedom in appointing faculty. If India were to do progressively well in science, the country can no longer treat colleges and universities where close to 95% of students go and 40% of the country's chemistry research is produced and which act as feeder institutions to research departments in leading institutions such as IISc and IITs with a step-motherly attitude. Professors at India's top institutions say that most students coming from state universities and colleges lack basic experimental skills and it takes a long time to train them. Apart from funding, many central universities (as well as state universities) suffer from retrograde personnel policies[48]. Universities can perform well if these problems are taken care of.

Government should boost the ease of doing science by making researchers less beholden to science administrators.[44] Funding agencies in India do not release funds in time and sanction money for maintenance/repairs for equipment. This has a downstream effect. Getting a research grant sanctioned means nothing until the researcher gets the money to procure the supplies and/or personnel and carry out the study. These problems affect especially those who have joined institutions which follow the tenure track system as they must prove themselves quickly. These are not insurmountable problems.[49]

Lack of jobs is perhaps the most serious problem. Thousands of PhD chemists in India face uncertainty, with each position attracting more than 250 applications. Finding under-employed Ph.Ds. after four postdocs is not uncommon, says a professor at IITM.[39] In contrast and against the global trend, in China the newly minted doctorates are gainfully employed in research. For example, in 2017 the unemployment rate of PhD graduates from Tsinghua University was just 2.4% and Peking University 3.18%.[50]

Aware of the need for large investments in R&D, Government has suggested that corporations give a part of their CSR funds to public research agencies and universities. CSR investment is welcome and can indeed be catalytic if transparent and flexible.
[https://twitter.com/thattai/status/1175974277054521346] One wonders why corporates would do that unless they have trust in these institutions. The Cipla Foundation's support to IISER Pune to build a state-of-the-art research laboratory for undergraduate students is a rare example that augurs well.[51]

To be able to carry out studies such as this one, one needs data related to research funding, undergraduate and doctoral degrees granted, manpower deployed by gender, participation by minorities, business R&D, etc. Such data should be regularly updated and made accessible in a structured way as is done in the US by the National Science Foundation. For now, what is available is R&D output data from sources such as Web of Science and Scopus. Comparative research about Indian science and Indian institutions is difficult due to lack of quality data about our research infrastructure and management.

**Acknowledgement**

We are grateful to Dr Venkat Nadella of Indian Institute of Science, Prof. M V Sangaranarayanan of Indian Institute of Technology Madras, and Prof. G Baskaran of the Institute of Mathematical Sciences for valuable suggestions.19

Fig. 1- Number of papers in chemistry from India and China in two different periods compared with that in selected advanced countries

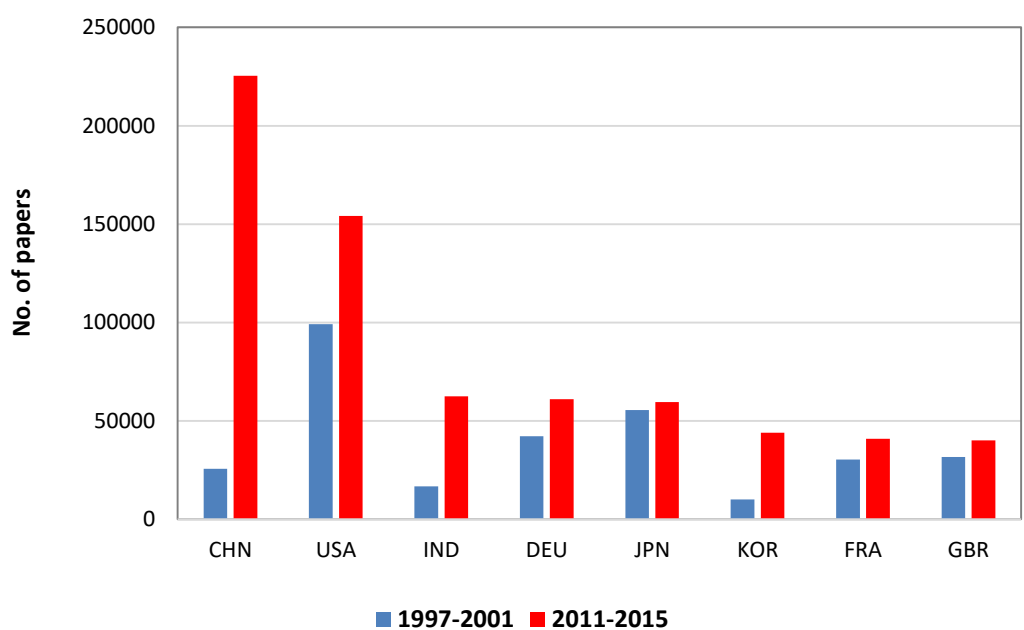

Fig. 2- Normalized citations per paper of World, India and China compared for eight subfields of chemistry

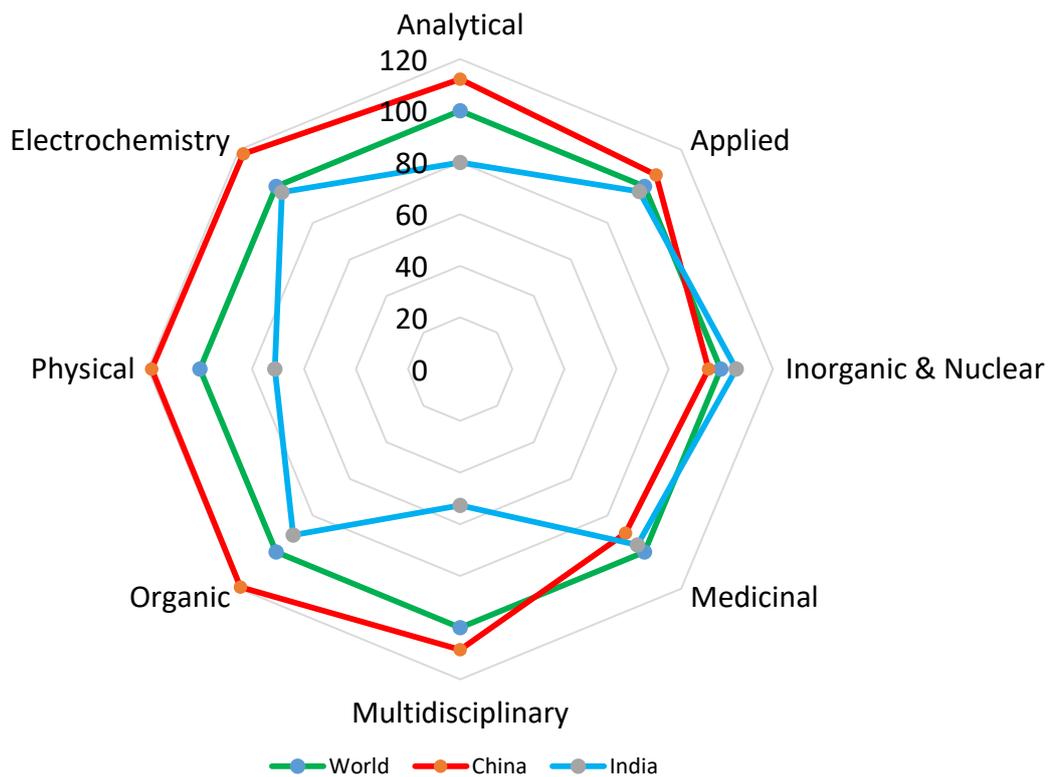

Fig. 3- Distribution of chemistry papers from Asian countries in the four quartiles of journals

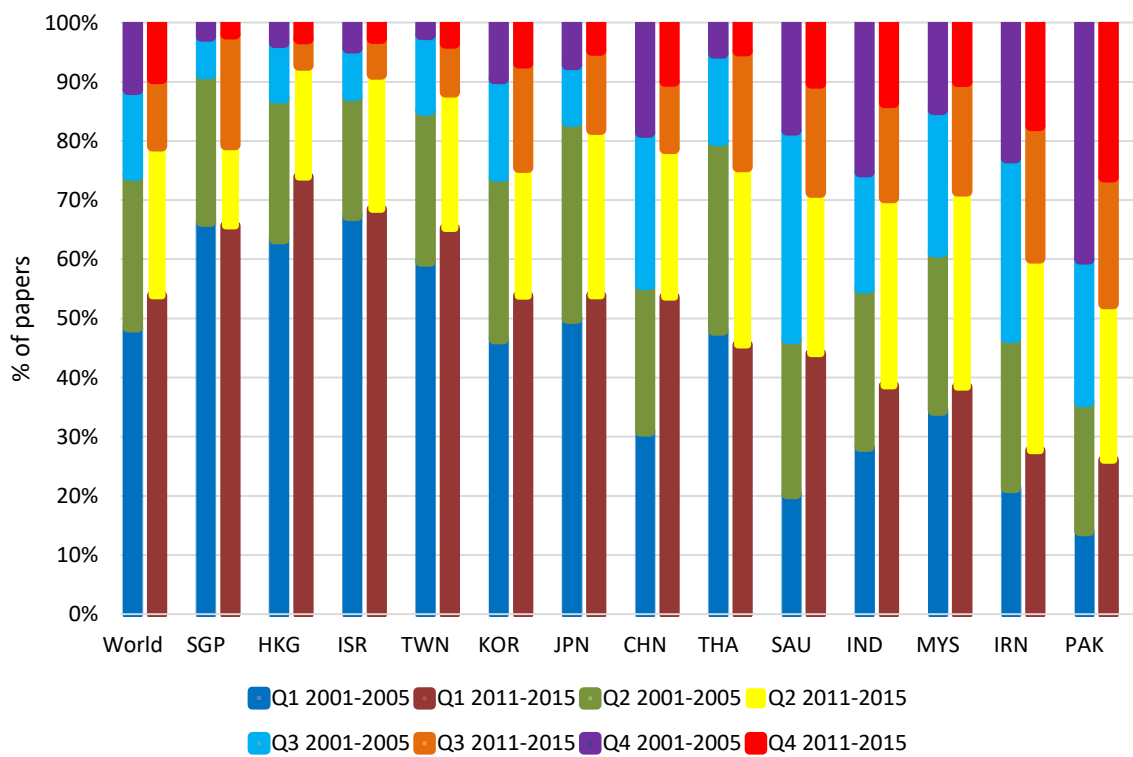

Fig. 4- Number of papers published from India and China during 2011-2015 and % of papers in Q1 journals.

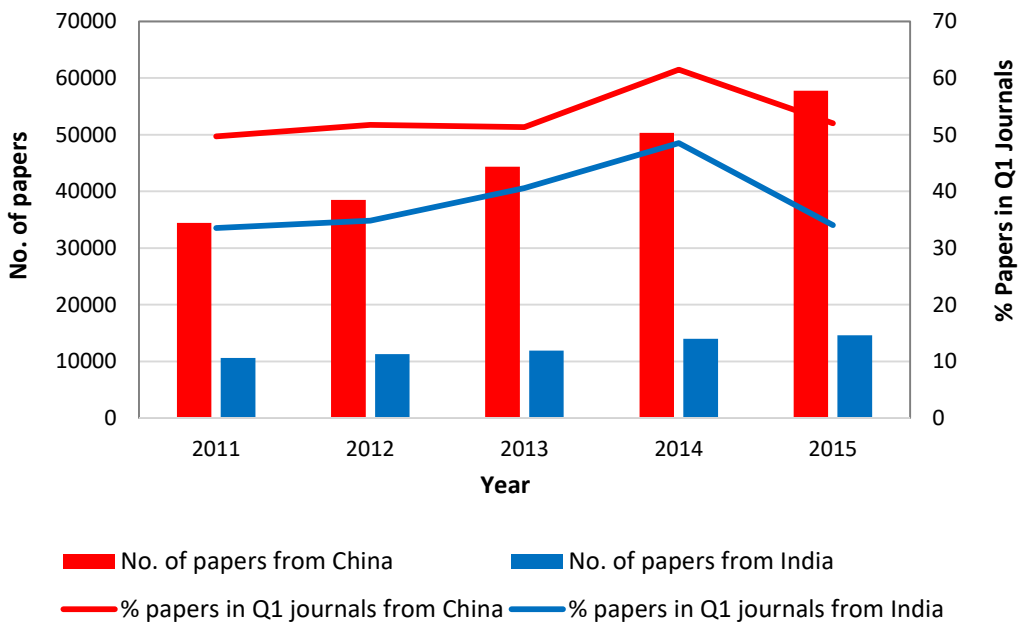

Fig. 5- Normalized % papers in Q1 journals of World, India and China compared for eight subfields of chemistry

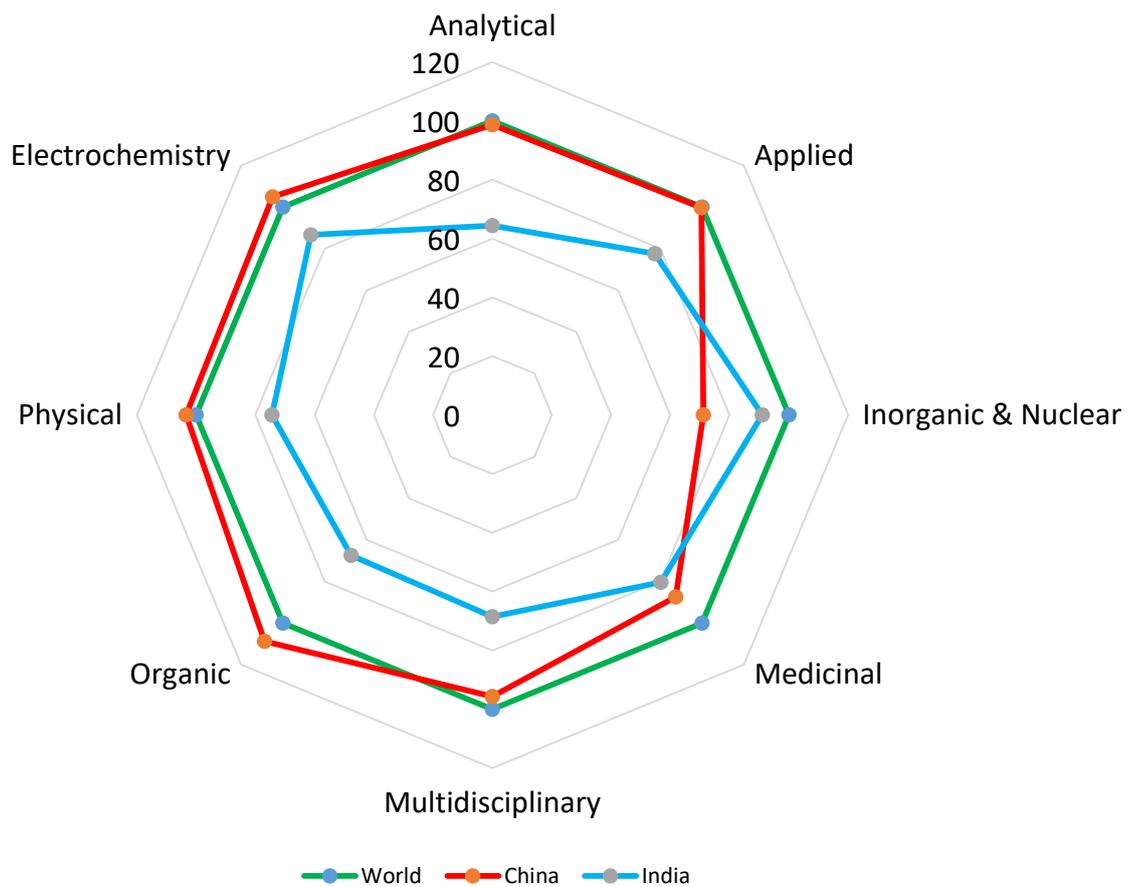

Fig. 6- Normalized % papers in top 1% of highly cited papers of World, India and China compared for eight subfields of chemistry

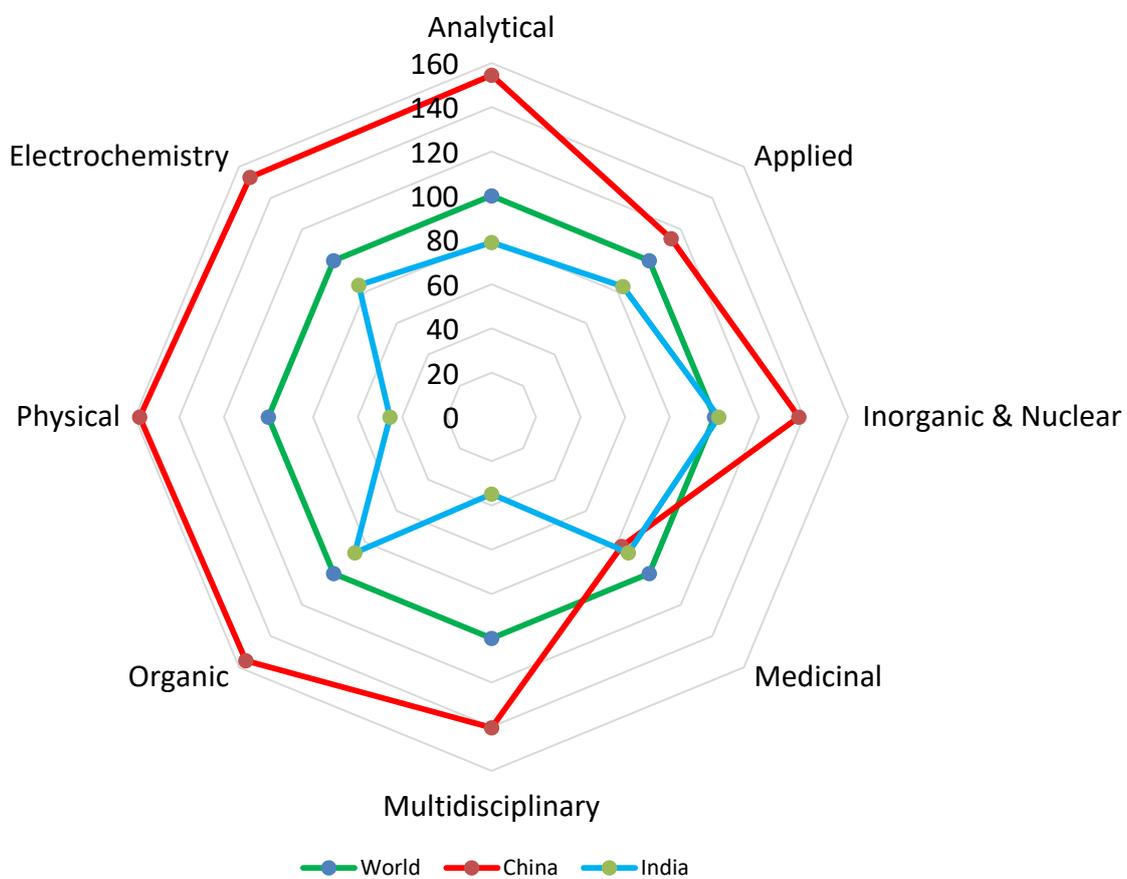

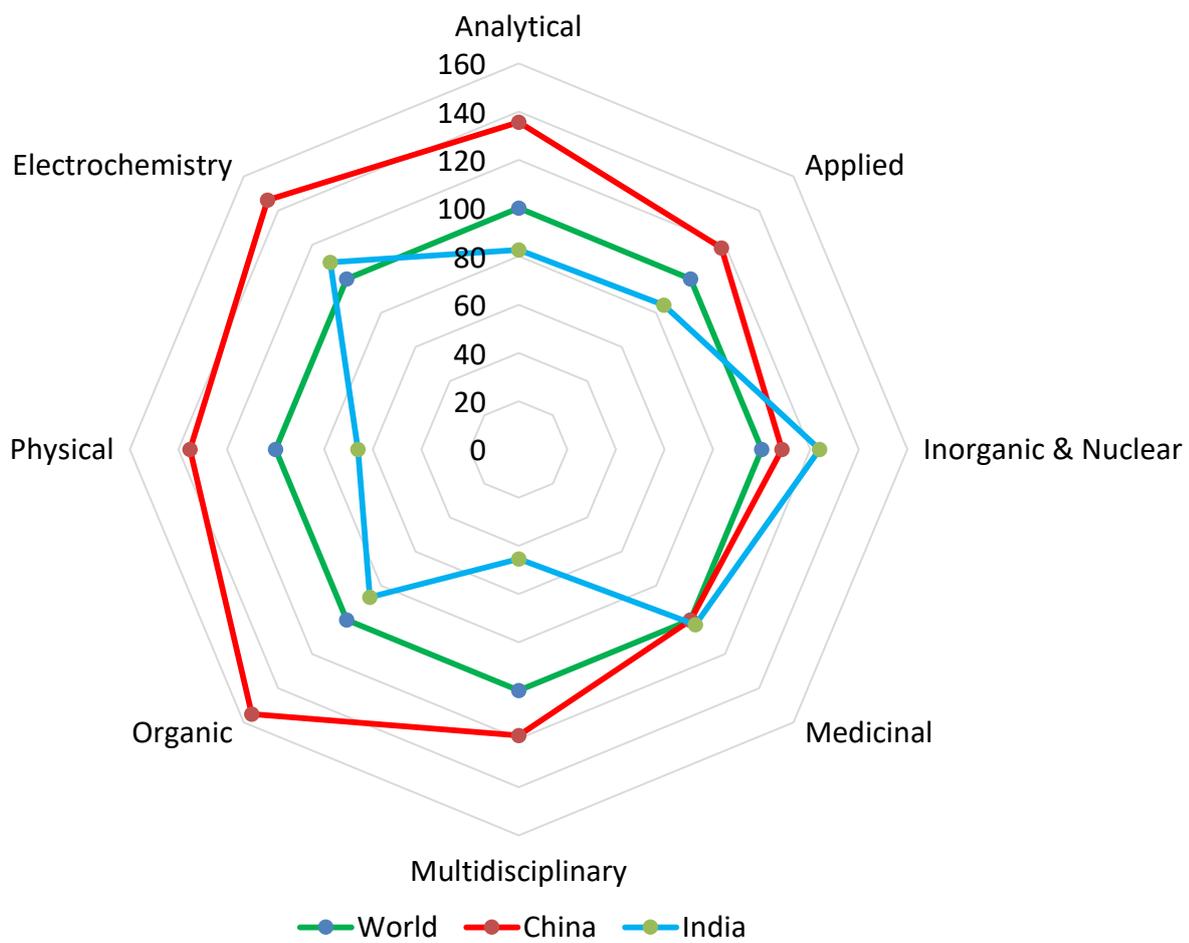

Fig. 7- Normalized % papers in top 10% of highly cited papers of World, India and China compared for eight subfields of chemistry

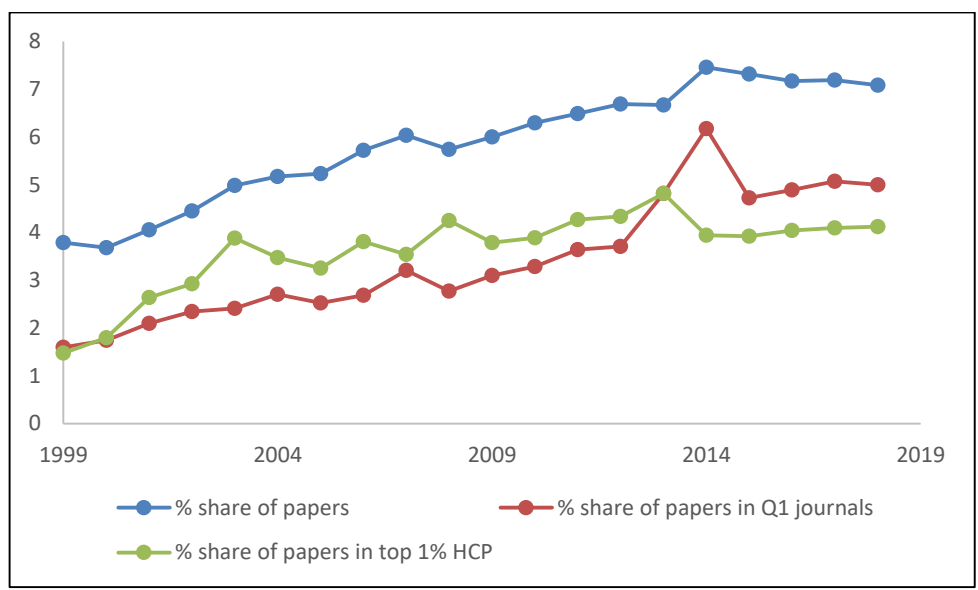

Fig. 8 India's % share of world's chemistry papers, % share of papers in Q1 journals and % share of papers in the top 1% HCP

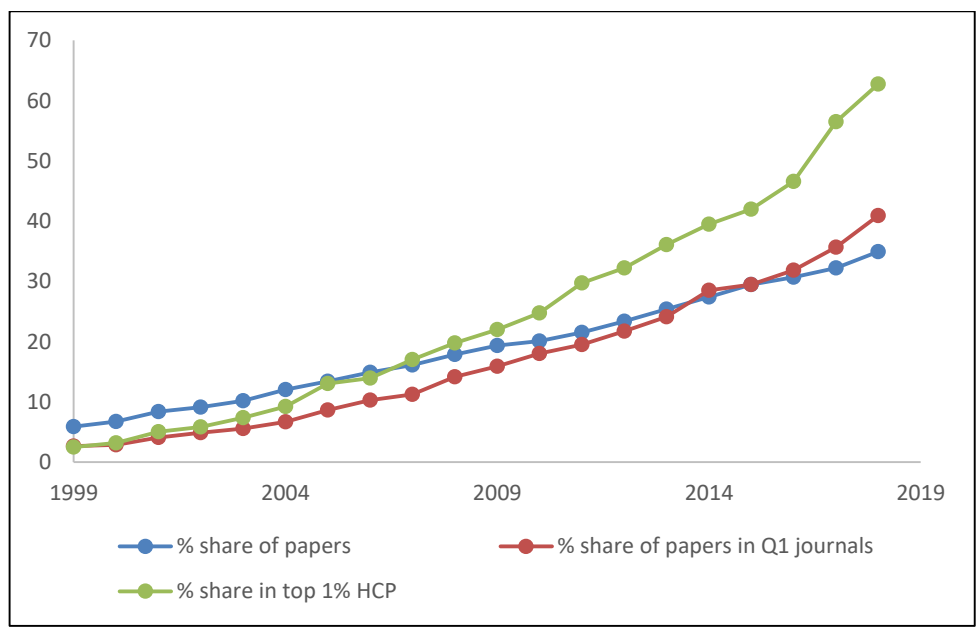

Fig. 9 China's % share of world's chemistry papers, % share of papers in Q1 journals and % share of papers in the top 1% HCP

Table 1: No. of papers from some groups of countries and some of their publication- and citation-based indicators

| Name | No. of papers | CPP | % Papers in Q1 journals | % Papers in top 1% | % Papers in Top 10% | % Int. Col. |
|---|---|---|---|---|---|---|
| Baseline* | 893985 | 20.67 | 54.10 | 1.46 | 13.08 | 21.75 |
| OECD Totals | 539096 | 22.90 | 61.54 | 1.57 | 13.88 | 32.69 |
| Asia Pacific Totals | 436626 | 20.97 | 52.07 | 1.66 | 14.12 | 20.87 |
| BRIC Totals | 337962 | 19.70 | 47.43 | 1.62 | 13.81 | 20.15 |
| EU-28 Totals | 271103 | 20.88 | 61.38 | 1.23 | 12.46 | 42.82 |
| EU-25 Totals | 263775 | 21.24 | 62.48 | 1.26 | 12.71 | 43.55 |
| EU-15 Totals | 233087 | 22.56 | 65.82 | 1.37 | 13.67 | 46.31 |
| Middle East Totals | 56784 | 16.14 | 37.02 | 1.18 | 10.78 | 37.17 |
| Latin America Totals | 34676 | 12.36 | 43.15 | 0.46 | 6.75 | 36.04 |
| ASEAN Totals | 23528 | 28.12 | 57.11 | 2.67 | 18.08 | 53.28 |
| Nordic Totals | 23173 | 22.14 | 66.82 | 1.50 | 13.84 | 61.60 |
| Africa Totals | 20327 | 12.34 | 34.08 | 0.54 | 7.59 | 60.43 |

*Baseline includes member countries of the groups listed

Table 2: Number of papers published from 42 countries which have published at least 4,000 papers during 2011-2015 and some of their publication- and citation-based indicators.

| No. | Name | No. of Papers | CPP | % Papers in Q1 Journals | % Papers in Top 1% | % Papers in Top 10% | % Int. Col. |
|---|---|---|---|---|---|---|---|
| | World | 900200 | 20.57 | 53.82 | 1.45 | 13.01 | 21.62 |
| 1 | China Mainland | 225407 | 23.12 | 53.61 | 2.09 | 16.86 | 19.12 |
| 2 | USA | 154093 | 30.47 | 69.11 | 2.65 | 19.40 | 37.85 |
| 3 | India | 62448 | 14.68 | 38.61 | 0.86 | 9.43 | 19.86 |
| 4 | Germany | 61030 | 24.77 | 67.22 | 1.60 | 14.51 | 51.04 |
| 5 | Japan | 59625 | 19.00 | 53.84 | 0.98 | 9.86 | 26.50 |
| 6 | South Korea | 44033 | 21.77 | 53.82 | 1.71 | 13.95 | 28.12 |
| 7 | France | 40978 | 21.04 | 65.71 | 1.04 | 11.83 | 57.19 |
| 8 | United Kingdom | 40021 | 26.51 | 68.86 | 1.87 | 16.20 | 56.55 |
| 9 | Spain | 36228 | 22.09 | 70.46 | 1.20 | 13.81 | 49.44 |
| 10 | Italy | 31440 | 20.31 | 61.77 | 1.28 | 13.39 | 47.50 |
| 11 | Russia | 30045 | 7.70 | 20.77 | 0.26 | 3.39 | 25.67 |
| 12 | Iran | 22858 | 13.71 | 27.65 | 0.79 | 9.52 | 17.57 |
| 13 | Canada | 22433 | 24.33 | 66.99 | 1.49 | 14.77 | 45.01 |
| 14 | Poland | 19744 | 11.54 | 38.76 | 0.32 | 5.59 | 33.51 |
| 15 | Australia | 18474 | 25.97 | 67.08 | 2.17 | 16.58 | 56.60 |
| 16 | Brazil | 18367 | 12.67 | 41.37 | 0.46 | 7.18 | 28.06 |
| 17 | Taiwan | 17582 | 19.26 | 65.24 | 1.19 | 11.48 | 24.42 |
| 18 | Switzerland | 13713 | 29.77 | 69.53 | 2.32 | 18.33 | 60.22 |
| 19 | Netherlands | 11230 | 29.28 | 75.21 | 2.27 | 18.08 | 60.80 |
| 20 | Turkey | 11072 | 12.62 | 37.38 | 0.88 | 8.78 | 25.67 |
| 21 | Singapore | 10229 | 45.34 | 78.50 | 4.86 | 28.93 | 54.05 |
| 22 | Belgium | 9962 | 22.56 | 67.09 | 1.26 | 15.05 | 66.79 |
| 23 | Sweden | 9955 | 23.63 | 68.88 | 1.79 | 14.81 | 64.58 |
| 24 | Saudi Arabia | 9427 | 24.36 | 44.08 | 2.95 | 16.91 | 82.42 |
| 25 | Czech Republic | 9080 | 15.35 | 47.04 | 0.81 | 8.48 | 47.43 |
| 26 | Portugal | 8964 | 19.23 | 57.86 | 1.03 | 12.70 | 53.67 |
| 27 | Egypt | 7980 | 12.70 | 28.71 | 0.51 | 7.94 | 56.64 |
| 28 | Hong Kong | 6965 | 33.74 | 73.91 | 3.33 | 23.86 | 24.51 |
| 29 | Malaysia | 6878 | 16.04 | 38.43 | 1.12 | 10.66 | 51.58 |
| 30 | Mexico | 6744 | 11.48 | 41.62 | 0.39 | 5.95 | 39.56 |
| 31 | Romania | 6526 | 8.90 | 27.26 | 0.23 | 4.26 | 37.14 |
| 32 | Denmark | 6488 | 24.35 | 69.28 | 1.82 | 15.49 | 62.36 |
| 33 | Austria | 6331 | 18.94 | 61.32 | 1.04 | 12.26 | 66.26 |
| 34 | Israel | 5221 | 25.09 | 68.45 | 1.59 | 14.84 | 50.18 |
| 35 | Ukraine | 4982 | 8.47 | 26.44 | 0.32 | 4.05 | 50.34 |
| 36 | Argentina | 4958 | 12.86 | 54.82 | 0.42 | 6.37 | 46.11 |
| 37 | Finland | 4828 | 19.96 | 64.31 | 0.87 | 12.59 | 64.21 |
| 38 | South Africa | 4719 | 14.52 | 42.53 | 0.59 | 9.22 | 51.56 |
| 39 | Greece | 4639 | 18.40 | 55.92 | 0.97 | 12.39 | 50.49 |
| 40 | Pakistan | 4516 | 11.12 | 26.04 | 0.80 | 7.79 | 53.12 |
| 41 | Hungary | 4452 | 13.61 | 46.29 | 0.36 | 7.08 | 52.79 |

| No. | Name | No. of Papers | CPP | % Papers in Q1 Journals | % Papers in Top 1% | % Papers in Top 10% | % Int. Col. |
|---|---|---|---|---|---|---|---|
| 42 | Thailand | 4197 | 14.44 | 45.53 | 0.98 | 9.65 | 41.46 |

151 other countries have published at least one paperer but less than 4,000

Table 3. CPP of papers published in Q1 and other than Q1 journals.*

| Category | Journals in quartiles | Q1 in all 5 years | Q1 in some years | Other than Q1 in all 5 years | Total |
|---|---|---|---|---|---|
| **World*** | No. of journals | 136 | 96 | 412 | 644 |
| | No. of papers | 393649 | 158713 | 355572 | 907934 |
| | Times cited | 13033218 | 2668622 | 2861135 | 18562975 |
| | CPP | 33.1 | 16.8 | 8.0 | 20.45 |
| **India** | No. of journals | 122 | 93 | 342 | 557 |
| | No. of papers | 17140 | 13278 | 32030 | 62448 |
| | Times cited | 450756 | 198662 | 267168 | 916586 |
| | CPP | 26.3 | 14.9 | 8.3 | 14.68 |
| **China** | No. of journals | 130 | 94 | 356 | 580 |
| | No. of papers | 97187 | 41323 | 86897 | 225407 |
| | Times cited | 3721753 | 797885 | 691462 | 5211100 |
| | CPP | 38.3 | 19.3 | 7.9 | 23.12 |

*Baseline data for 193 countries, and the difference in the baseline total is due to changes in the inCites data policy during our data collection.

Table 4: State and Central Universities that have published at least 500 papers*

| No. | Name | Type of univ | No. of Papers | CPP | % Papers in Q1 Journals | % Papers in Top 1% | % Papers in Top 10% | % Int. Col. |
|---|---|---|---|---|---|---|---|---|
|  | Baseline |  | 25119 | 13.60 | 31.22 | 0.73 | 9.00 | 20.18 |
| 1 | Jadavpur University | State | 1278 | 14.69 | 35.05 | 0.86 | 11.97 | 24.49 |
| 2 | University of Delhi | Central | 1271 | 17.96 | 43.35 | 0.63 | 11.64 | 25.81 |
| 3 | Banaras Hindu University | Central | 1230 | 17.58 | 40.73 | 0.98 | 12.11 | 23.50 |
| 4 | University of Hyderabad | Central | 1088 | 16.05 | 48.44 | 0.92 | 9.93 | 15.99 |
| 5 | University of Calcutta | State | 1004 | 14.04 | 36.16 | 0.70 | 9.96 | 28.29 |
| 6 | Institute of Chemical Technology - Mumbai | State | 745 | 16.72 | 38.26 | 0.40 | 13.69 | 14.90 |
| 7 | Anna University | State | 718 | 11.79 | 35.79 | 0.42 | 8.22 | 24.79 |
| 8 | Aligarh Muslim University | Central | 663 | 13.33 | 26.24 | 0.75 | 9.05 | 33.63 |
| 9 | Panjab University | State | 660 | 13.70 | 31.36 | 0.30 | 8.64 | 27.58 |
| 10 | Guru Nanak Dev University | State | 636 | 19.53 | 49.21 | 2.36 | 14.47 | 23.90 |
| 11 | Annamalai University | State | 575 | 9.80 | 17.22 | 0.52 | 4.35 | 16.87 |
| 12 | University of Pune | State | 544 | 13.45 | 40.44 | 0.37 | 7.72 | 26.65 |
| 13 | University of Madras | State | 540 | 13.25 | 32.04 | 1.30 | 7.96 | 11.48 |
| 14 | Madurai Kamaraj University | State | 503 | 19.06 | 35.79 | 1.39 | 11.73 | 32.41 |

*67 other state and central universities have published at least 100 papers but less than 500.

Table 5. Private Institutions that have published at least 100 papers

| No. | Name | No. of Papers | CPP | % Papers in Q1 Journals | % Papers in Top 1% | % Papers in Top 10% | % Int. Col. |
|---|---|---|---|---|---|---|---|
|  | Baseline | 2648 | 13.81 | 32.14 | 0.98 | 9.78 | 23.49 |
| 1 | Vellore Institute of Technology | 684 | 10.85 | 21.64 | 0.44 | 7.60 | 20.91 |
| 2 | Birla Institute of Technology and Science Pilani (BITS Pilani) | 532 | 14.21 | 34.77 | 0.56 | 9.02 | 26.88 |
| 3 | Thapar University | 251 | 11.39 | 29.88 | 0.00 | 4.78 | 12.35 |
| 4 | Birla Institute of Technology Mesra | 218 | 13.77 | 35.32 | 0.46 | 10.09 | 24.77 |
| 5 | Amrita Vishwa Vidyapeetham University | 188 | 31.21 | 57.98 | 4.79 | 23.40 | 34.57 |
| 6 | Manipal University | 181 | 10.72 | 34.81 | 0.00 | 8.29 | 20.44 |
| 7 | Amity University | 162 | 14.21 | 38.27 | 1.23 | 10.49 | 26.54 |
| 8 | Loyola College - India | 144 | 13.58 | 33.33 | 2.08 | 13.19 | 20.83 |
| 9 | SRM Institute of Science & Technology | 140 | 17.42 | 37.86 | 2.14 | 15.00 | 22.14 |
| 10 | Gandhi Institute of Technology & Management (GITAM) | 118 | 6.07 | 11.02 | 0.85 | 3.39 | 14.41 |
| 11 | Karunya Institute of Technology & Sciences | 111 | 14.31 | 27.03 | 1.80 | 11.71 | 37.84 |

Table 6. MHRD Institutions that have published at least 300 papers*

| No. | Name | No. of Papers | CPP | % Papers in Q1 Journals | % Papers in Top 1% | % Papers in Top 10% | % Int. Col. |
|---|---|---|---|---|---|---|---|
| | Baseline | 13530 | 18.88 | 54.08 | 1.48 | 13.22 | 21.05 |
| 1 | Indian Institute of Science (IISC) - Bangalore | 2079 | 22.75 | 62.82 | 2.12 | 15.68 | 19.34 |
| 2 | Indian Institute of Technology (IIT) - Bombay | 1397 | 17.84 | 60.49 | 0.86 | 11.81 | 26.91 |
| 3 | Indian Institute of Technology (IIT) - Kharagpur | 1331 | 17.73 | 54.47 | 1.05 | 9.32 | 16.98 |
| 4 | Indian Institute of Technology (IIT) - Madras | 1203 | 18.55 | 54.78 | 1.08 | 12.14 | 17.46 |
| 5 | Indian Institute of Technology (IIT) - Kanpur | 1130 | 16.30 | 58.41 | 1.24 | 12.12 | 25.66 |
| 6 | Indian Institute of Technology (IIT) - Guwahati | 860 | 17.09 | 53.02 | 0.93 | 13.14 | 9.53 |
| 7 | Indian Institute of Technology (IIT) - Delhi | 831 | 17.42 | 54.51 | 1.32 | 11.43 | 23.47 |
| 8 | Indian Institute of Technology (IIT) - Roorkee | 811 | 25.39 | 43.65 | 4.19 | 18.25 | 34.03 |
| 9 | Indian Institute of Science Education & Research (IISER) - Kolkata | 419 | 17.08 | 63.72 | 0.48 | 10.26 | 21.00 |
| 10 | Indian Institute of Science Education & Research (IISER) Pune | 402 | 24.00 | 75.87 | 2.99 | 15.92 | 14.18 |
| 11 | Indian Institute of Technology (IIT BHU) - Varanasi | 401 | 14.71 | 41.65 | 0.50 | 10.47 | 25.19 |
| 12 | Indian Institute of Engineering Science Technology Shibpur (IIEST) | 335 | 19.50 | 45.67 | 1.49 | 16.72 | 24.18 |

Table 7. CSIR Institutions publishing at least 200 papers*†

| No. | Name | No. of Papers | CPP | % Papers in Q1 Journals | % Papers in Top 1% | % Papers in Top 10% | % Int. Col. |
|---|---|---|---|---|---|---|---|
|  | Baseline | 10250 | 16.92 | 47.74 | 0.9 | 10.68 | 15.66 |
| 1 | Indian Institute of Chemical Technology | 2553 | 14.46 | 33.65 | 0.43 | 8.42 | 15.94 |
| 2 | National Chemical Laboratory | 1605 | 20.39 | 60.19 | 1.06 | 12.4 | 15.89 |
| 3 | Central Drug Research Institute | 737 | 13.94 | 44.78 | 0.54 | 8.68 | 6.51 |
| 4 | Central Salt & Marine Chemical Research Institute | 612 | 18.99 | 53.27 | 1.63 | 11.93 | 12.58 |
| 5 | National Physical Laboratory - India | 572 | 25.39 | 66.08 | 1.92 | 18.18 | 26.75 |
| 6 | Central Leather Research Institute | 564 | 13.99 | 42.25 | 0.71 | 9.04 | 14.72 |
| 7 | Academy of Scientific and Innovative Research (AcSIR)‡ | 524 | 15.9 | 58.02 | 1.72 | 11.64 | 7.82 |
| 8 | Central Electrochemical Research Institute | 503 | 19.82 | 60.44 | 1.39 | 12.92 | 23.66 |
| 9 | National Institute Interdisciplinary Science & Technology | 426 | 21.94 | 61.03 | 0.7 | 12.21 | 21.13 |
| 10 | Indian Institute of Integrative Medicine (IIIM), Jammu | 385 | 14.84 | 43.12 | 1.04 | 10.13 | 5.97 |
| 11 | Indian Institute of Chemical Biology | 325 | 12.19 | 43.08 | 0.62 | 8.31 | 9.54 |
| 12 | Central Institute of Medicinal & Aromatic Plants | 234 | 7.44 | 17.52 | 0 | 3.85 | 6.84 |
| 13 | North East Institute of Science and Technology, Jorhat | 216 | 17.7 | 42.59 | 1.85 | 8.8 | 8.8 |
| 14 | Institute of Minerals and Materials Technology | 216 | 20.5 | 62.5 | 1.85 | 14.81 | 12.5 |
| 15 | Central Glass & Ceramic Research Institute | 211 | 16.7 | 62.09 | 0.47 | 16.11 | 11.37 |
| 16 | Indian Institute of Petroleum | 203 | 18.86 | 59.61 | 0.49 | 11.33 | 21.18 |

*4 other CSIR institutions have published at least 100 papers but less than 200.

†The number of papers from each laboratory was assigned after manually checking the bylines of all the papers, because there were variations in the way *InCites* had assigned them.

‡AcSIR is an artifact. In reality, the science is performed by the same scientists and their collaborators in their own laboratories which are part of CSIR. But because AcSIR is given in the byline we have taken note of it.

Table 8. DAE Institutions that have published at least 100 papers in chemistry

| No. | Name | No. of Papers | CPP | % Papers in Q1 Journals | % Papers in Top 1% | % Papers in Top 10% | % Int. Col. |
|---|---|---|---|---|---|---|---|
|  | Baseline | 3582 | 13.23 | 51.59 | 0.67 | 8.24 | 17.17 |
| 1 | Bhabha Atomic Research Center | 2247 | 12.76 | 48.64 | 0.62 | 7.65 | 15.13 |
| 2 | Indira Gandhi Centre for Atomic Research | 493 | 10.03 | 47.67 | 0.41 | 6.29 | 10.34 |
| 3 | Tata Institute of Fundamental Research | 288 | 17.27 | 67.71 | 1.04 | 12.50 | 38.89 |
| 4 | Saha Institute of Nuclear Physics | 286 | 13.57 | 50.70 | 1.40 | 9.44 | 11.54 |
| 5 | National Institute of Science Education & Research (NISER) | 198 | 21.08 | 65.66 | 1.01 | 13.13 | 32.83 |
| 6 | Raja Ramanna Centre for Advanced Technology | 127 | 10.84 | 63.78 | 0.00 | 7.09 | 14.96 |
| 7 | Institute of Physics Bhubaneswar (IOPB) | 106 | 16.07 | 77.36 | 0.94 | 11.32 | 26.42 |

Table 9. DST Institutions that have published at least 100 papers in chemistry

| No. | Name | No. of Papers | CPP | % Papers in Q1 Journals | % Papers in Top 1% | % Papers in Top 10% | % Int. Col. |
|---|---|---|---|---|---|---|---|
|  | Baseline | 2861 | 21.17 | 60.92 | 1.29 | 13.14 | 20.34 |
| 1 | Indian Association for the Cultivation of Science (IACS) - Jadavpur | 1427 | 20.30 | 64.19 | 0.84 | 12.75 | 16.68 |
| 2 | Jawaharlal Nehru Center for Advanced Scientific Research | 682 | 29.00 | 70.82 | 2.35 | 19.21 | 20.82 |
| 3 | SN Bose National Centre for Basic Science | 205 | 17.11 | 55.12 | 1.46 | 9.27 | 29.27 |
| 4 | Bose Institute | 167 | 12.10 | 33.53 | 0.60 | 5.39 | 19.76 |
| 5 | Raman Research Institute | 167 | 15.06 | 40.72 | 0.60 | 5.39 | 29.94 |
| 6 | Centre for Nano & Soft Matter Sciences | 154 | 13.82 | 36.36 | 1.30 | 6.49 | 35.71 |
| 7 | Sree Chitra Tirunal Institute for Medical Sciences Technology | 100 | 20.87 | 77.00 | 2.00 | 17.00 | 10.00 |

Table 10. Other Institutions that have published at least 100 papers in chemistry

| No. | Name | Type* | No. of Papers | CPP | % Papers in Q1 Journals | % Papers in Top 1% | % Papers in Top 10% | % Int. Col. |
|---|---|---|---|---|---|---|---|---|
| | Baseline | | 1841 | 12.56 | 43.35 | 0.43 | 8.37 | 14.12 |
| 1 | National Institute of Pharmaceutical Education & Research (NIPER) | MCF | 649 | 14.92 | 38.21 | 0.62 | 10.48 | 9.86 |
| 2 | Indian Council of Agricultural Research (ICAR) | ICAR | 387 | 9.48 | 32.56 | 0.52 | 7.24 | 12.14 |
| 3 | Department of Biotechnology (DBT) India | DBT | 205 | 13.58 | 48.78 | 0.49 | 9.27 | 20.00 |
| 4 | UGC DAE Consortium for Scientific Research | UGC | 196 | 11.27 | 57.14 | 0.51 | 6.63 | 19.39 |
| 5 | Advanced Centre Research in High Energy Materials (ACRHEM) | DRDO | 151 | 13.74 | 40.40 | 0.00 | 5.30 | 16.56 |
| 6 | Inter-University Accelerator Centre | UGC | 140 | 9.88 | 71.43 | 0.00 | 7.14 | 31.43 |
| 7 | Defense Research & Development Establishment (DRDE) | DRDO | 134 | 11.96 | 46.27 | 0.00 | 8.96 | 3.73 |

*Institution type was assigned based on the byline information provided in *InCites*